\documentclass[pre,aps,twocolumn,superscriptaddress,nofootinbib]{revtex4-2}
\usepackage{graphicx}
\usepackage{amssymb,latexsym,amsthm,amsmath}    
\usepackage{mathtools}
\usepackage{float} 
\usepackage[draft]{changes}
\usepackage{hyperref}
\usepackage{color}
\usepackage{hhline}
\usepackage{colortbl}
\usepackage{ulem} 
\usepackage{xcolor}

\newcommand{\mean}[1]{\left\langle #1 \right\rangle}

\begin{document}

\title{Cooperation, satisfaction, and rationality in social games on complex networks with aspiration-driven players}
\author{M. Aguilar-Janita}
\email[Corresponding author:]{miguelaj@unex.es}
\affiliation{Departamento de F\'{\i}sica,
  Universidad de Extremadura, 06006 Badajoz,
  Spain}
\author{N. Khalil}%
\affiliation{Departamento de Física y Matemáticas \& Grupo Interdisciplinar de Sistemas Complejos, Universidad de Alcalá, 28805 Alcalá de Henares, Spain}
\author{I. Leyva}
\affiliation{Complex Systems Group \& Grupo Interdisciplinar de Sistemas Complejos, Universidad Rey Juan Carlos, 28933 M\'ostoles, Spain}
\affiliation{Center for Biomedical Technology, Universidad Polit\'ecnica de Madrid, 28223 Pozuelo de Alarc\'on, Spain}
\author{I. Sendi\~na-Nadal}
\affiliation{Complex Systems Group \& Grupo Interdisciplinar de Sistemas Complejos, Universidad Rey Juan Carlos, 28933 M\'ostoles, Spain}
\affiliation{Center for Biomedical Technology, Universidad Polit\'ecnica de Madrid, 28223 Pozuelo de Alarc\'on, Spain}

\begin{abstract}
A network model based on players' aspirations is proposed and analyzed theoretically and numerically within the framework of evolutionary game theory. In this model, players decide whether to cooperate or defect by comparing their payoffs from pairwise games with their neighbors, driven by a common aspiration level. The model also incorporates a degree of irrationality through an effective temperature in the Fermi function. The level of cooperation in the system is fundamentally influenced by two social attributes: satisfaction, defined as the fraction of players whose payoffs exceed the aspiration level, and the degree of rationality in decision-making. Rational players tend to maintain their initial strategies for sufficiently low aspiration levels, while irrational agents promote a state of perfect coexistence, resulting in half of the agents cooperating.
The transition between these two behaviors can be critical, often leading to abrupt changes in cooperation levels. When the aspiration level is high, all players become dissatisfied, regardless of the effective temperature. Intermediate aspiration levels result in diverse behaviors, including sudden transitions for rational agents and a non-monotonic relationship between cooperation and increased irrationality. The study also carefully examines the effects of the interaction structure, initial conditions, and the strategy update rule (asynchronous versus synchronous). Special attention is given to the prisoner's dilemma, where significant cooperation levels can be achieved in a structured environment, with moderate aspiration and high rationality settings, and following a synchronous strategy updating scheme.

\end{abstract}

\maketitle

\section{Introduction}
\label{sec:intro}
When we attempt to understand the behavior of biological or human systems through mathematical models, we encounter problems similar to those when dealing with almost any other complex system: What are the dynamics of one or a few individuals? How do interactions take place, and how do they influence individual dynamics? What role, if any, do internal and external noise or forces play? How do we combine all the microscopic features to explain the macroscopic/observable behavior? Etc.

Primitive models typically focus on only a subset of the previous questions, often neglecting others. For instance, replicator dynamics \cite{schuster1983replicator, roca2009evolutionary} used in evolutionary game theory operates on a deterministic system of equations that describes the fraction of individuals possessing a given trait, assuming it fully captures the system's behavior. Later developments incorporated factors such as noise  \cite{cabrales1993stochastic} and structured interactions \cite{lieberman2005evolutionary}, among others. These elements are now considered essential for any sufficiently realistic system description.

In this work, we model social behavior by examining only the most relevant aspects. Our first simplification is to consider individuals as existing in one of two states, cooperator (C) or defector (D), without assigning them any specific meaning. This approach aligns with the understanding that two-state models are both simple and powerful tools for studying various real-world characteristics of physical and biological systems. Examples of such two-state models include Ising-like models \cite{galam1982sociophysics,galam2010ising}, birth-death processes in biology and ecology \cite{marro2005nonequilibrium,klemm2020altruism}, and the Voter Model in social systems \cite{castellano2009statistical,aguilar2024polarization}, among others.

We assume that an agent's state evolves according to the principles of the evolutionary game theory within a structured and fixed-time environment \cite{roca2009evolutionary, tanimoto2015fundamentals}. Essentially, an agent decides whether to change her strategy by analyzing the payoff obtained at each game round with her neighbors, who keep their strategy constant. A dyadic game models this interaction \cite{von1944theory, maynard1976evolution}, which assigns a payoff to each player. The payoff often depends on the roles of the two players (cooperating or defecting) and is typically expressed as a linear function of the game parameters.

The potential changes in strategy stem from analyzing this payoff. In the literature,  decision-making processes are modeled through various mechanisms, with imitation-driven \cite{roca2009evolutionary,khalil2023deterministic} and aspiration-driven \cite{lim2018satisfied} approaches being the most relevant to our study. Here, we assume that the dynamics is influenced by a common intrinsic aspiration shared among the players, eventually modulated by some degree of irrationality.

When imitation is the sole mechanism for evolution, an agent compares her payoff to a neighbor's and is prone to change her strategy if her payoff is lower. A key achievement of this family of models is their theoretical explanation for the prevalence of cooperative and altruistic behavior in natural and social systems \cite{nowak_n92b,szabo_pr07}.

    In aspiration-driven evolution~\cite{lim2018satisfied, du_j_sr15,zhou_l_nc21,du_j_jrsi14,cimini2015,platkowski_aml09,posch_prslb99,perc_pone10}, agents assess their payoffs based on an intrinsic feature known as {\it aspiration} or {\it mood} to determine whether to change strategies independent of other players' actions. This mechanism is relevant for either humans \cite{roca_pnas11} or animals \cite{gruter_bes11}. Our study uses a model similar to those presented in \cite{du_j_sr15, lim2018satisfied, zhou_l_nc21}; unlike those studies, we do not restrict our analysis to weak selection with highly irrational agents. Instead,  we consider all levels of rationality, including fully rational agents. Notably, non-universal dynamic properties can emerge in this context, leading to abrupt cooperation-level transitions \cite{Zhuk2021, khalil2023deterministic}.

Another source of interest in aspiration-driven evolution arises from behavioral laboratory experiments conducted with humans. These experiments suggest that the level of cooperation is not influenced by the network topology \cite{grujic_pone10,grujic_srep12,gracia-lazaro_pnas12,gracia-lazaro_srep12}. It has been proposed \cite{cimini2015} that this behavior is better modeled by dynamics that do not take into account the neighbors' payoffs, such as the aspiration dynamics. Reference ~\cite{du_j_sr15} analytically demonstrates that, under weak selection, the favored strategy is not sensitive to the underlying (regular) graph and remains the same as in a well-mixed population. This finding motivates us to extend our study \cite{aguilar2024cooperation} to encompass complex network topologies. We anticipate that the results reported in Ref.~\cite{du_j_sr15} no longer hold for moderately to highly rational agents.

An important distinction exists between imitation and aspiration mechanisms that is relevant to the theoretical description of our systems. In imitation, reciprocity influences both the payoffs and the copying process. In contrast, in aspiration-driven mechanisms neighbors only impact the payoffs, which likely results in weaker reciprocity effects. This property could aid in developing an approximate theory that has yet to be explored.

When implementing our model, a key consideration is whether to use synchronous or asynchronous strategy update rules. Although asynchronous updates may appear more natural in systems that operate without external control, synchronous updates are more frequently used in experiments involving humans. As demonstrated later, these two approaches can lead to different outcomes, which may often be surprising or counterintuitive.

When implementing our model, a key consideration is whether to use synchronous or asynchronous strategy updating rules. While the latter might seem more natural in systems that operate without external control, the former is more frequently used in experiments involving humans. As demonstrated later, these two approaches can lead to different outcomes, which may often be surprising or counterintuitive.

In this work, we re-evaluate the model presented in \cite{aguilar2024cooperation}, expanding our analysis beyond the mean-field approach. We discovered in \cite{aguilar2024cooperation} that an appropriate rescaling simplifies the relevant parameters, allowing for a more comprehensive exploration of the system's behavior. As a result, we identified various states for rational agents, including absorbing states, consensus, and coexistence, and examined how these states are affected as irrationality increases. Notably, in the context of the prisoner's dilemma, cooperation is highly dependent on initial conditions when agents are rational and have low aspirations. As irrationality and aspirations rise, this dependence weakens, sometimes abruptly. Furthermore, an additional increase in irrationality leads the cooperation fraction to approach \(1/2\). The present study aims to demonstrate how these dynamics shift within structured populations.

The remainder of this paper is organized as follows: In the next section, Sec. \ref{sec:model}, we will define the model, rescale it, reduce the number of effective parameters, and derive an approximate theoretical description from a more complex, though less tractable, one. In Sec. \ref{sec:rationalagents}, we focus on rational agents, identifying and characterizing different stationary states similar to those already found in the mean-field analysis. We provide estimates for the range of scaling parameters for which these states may appear and pay particular attention to the role of the underlying network in potential abrupt changes in the cooperation fraction as parameters and initial conditions vary.

In Sec. \ref{sec:irrationalagents}, we explore the case of irrational agents, analyzing how an increase in the effective temperature influences the various steady states and the critical behavior of the system. Section \ref{sec:prisonersdilemma} focuses on the prisoner's dilemma game, where we provide extensive numerical simulations alongside a theoretical analysis based on the findings from the previous sections. The results reveal a consistent agreement between theory and simulations regarding the level of cooperation and its fluctuations as the game parameters, aspiration, and irrationality are adjusted. The final section presents the main conclusions.

\section{Model}
\label{sec:model}

We consider a connected and undirected network \(\Sigma\) with \(N\) nodes, each one representing an agent with two possible strategies: cooperation (C) or defection (D). Nodes' connectivity is given by the adjacency matrix \(\mathcal A\) so that:
\begin{equation}
  \mathcal{V}_\mu=\{\nu\in\Sigma \, :\, \mathcal A_{\mu,\nu}=1\}  
\end{equation} 
is the neighbourhood of node \(\mu\) and
\begin{equation}
  k_\mu=\sum_{\nu\in\mathcal V_\mu}A_{\mu,\nu}  
\end{equation}
is the degree of node \(\mu\).

A node is always occupied by an agent with a well-defined strategy. Hence, at a given time, the state of the system is completely specified by the vector
\begin{equation}
  \textbf S=(c_1,\dots,c_N)
\end{equation}
where \(c_\mu=1\) if node \(\mu\) has a cooperation strategy and \(c_\mu=0\) otherwise.

We model the evolution of agents using two complementary strategy update rules: asynchronous and synchronous. In both cases, the dynamics involves several steps:
\begin{itemize}
\item [(i)] The network $\mathcal A$, the initial configuration (or microstate) $\textbf S_0$, and the game parameters are set.
\item [(ii)] In each round, every agent \(\mu\) plays the same game with her neighbors and receives a payoff \(g_\mu\). This payoff depends on the strategies of \(\mu\) and her neighbors, as well as the game parameters: the reward for mutual cooperation (\(R\)), the sucker’s payoff (\(S\)), the temptation to defect (\(T\)), and the punishment for mutual defection (\(P\)). They payoff is computed as:
  \begin{equation}
    \label{eq:gsigma}
    g_\mu=c_\mu\sum_{\nu\in \mathcal{V}_\mu}(Rc_\nu+Sd_\nu)+d_\mu\sum_{\nu\in \mathcal{V}_\mu}(Tc_\nu+Pd_\nu),
  \end{equation}
  where
  \begin{equation}
    d_\mu=1-c_\mu.
  \end{equation}
\item[(iii)] Depending on the selected strategy updating method, only one randomly chosen agent (asynchronous rule) or all agents (synchronous) can change, at each round, their strategy with probability:
\begin{equation}
  \label{eq:muupdate}
  p_\mu=\frac{1}{1+\exp\left(\frac{s_\mu}{\theta }\right)},
\end{equation}
with
\begin{eqnarray}
  \label{eq:smu}
  \nonumber s_\mu=\left(\frac{g_\mu}{k_\mu}-m\right)\left[c_\mu \max(|R-m|,|S-m|)\right. && \\ \left.  +d_\mu\max(|T-m|,|P-m|)\right]^{-1}. && 
\end{eqnarray}
The parameter $m\in \mathbb R$ stands for the {\it aspiration or mood}, the same for all agents, and $\theta\ge 0$ is an effective temperature, a mean of the level of irrationality of choices.
\item[(iv)] The time $t$ is updated to \(t\to t+t_0/N\) for asynchronous updates, or to \(t\to t+t_0\) for synchronous ones, where \(t_0\) represents a unit of time. 
\end{itemize}
Steps (ii) to (iv) are repeated until a desired time {$t_{\rm total}$} is reached. 

This model simplifies to the one analyzed in Ref. \cite{aguilar2024cooperation} for fully connected networks. A key difference from imitation-driven models, such as in Ref. \cite{khalil2023deterministic}, is that the probability \(p_\mu\) of changing strategies in Eq.~\eqref{eq:muupdate} depends solely on the agent's payoff rather than on the payoffs of their neighbors. Additionally, the transition probability \(p_\mu\) is also influenced by the effective temperature \(\theta\) and the aspiration or mood \(m\). When \(\theta=0\), agents are fully rational, sticking to a strategy as long as their payoffs meet or exceed their expectations \(m\). As \(\theta\) increases, agents' behavior trends towards irrationality. Although the model could accommodate heterogeneous effective temperatures, aspiration parameters, and game parameters among agents, such complexities are not included in this analysis. Notably, the aspiration level \(m\) can be negative, which enables agents to resist changing strategies even when facing negative payoffs.

\subsection{Parametric re-scaling}

As it is apparent from Eq.~\eqref{eq:muupdate}, the dependence of the model on the many parameters occurs through \(s_\mu\), given by Eq.~\eqref{eq:smu}, and the effective temperature \(\theta\). However, \(s_\mu\) can be rewritten as 
\begin{eqnarray}
  \nonumber 
  s_\mu&=&\frac{c_\mu}{k_\mu\max(1,|\sigma|)}\sum_{\nu\in\mathcal V_\mu}(\kappa_{c} c_\nu+\sigma d_\nu)\\ &&+\frac{d_\mu}{k_\mu\max(1,|\tau|)}\sum_{\nu\in\mathcal V_\mu}(\tau c_\nu+\kappa_{d} d_\nu),
\end{eqnarray}
with four new game-aspiration parameters  
\begin{eqnarray}
 \sigma=\frac{S-m}{|R-m|}  &,& \hspace{0.7cm} \tau=\frac{T-m}{|P-m|} \\
  \kappa_{c} =\frac{R-m}{|R-m|} &,&  \hspace{0.7cm} \kappa_{d} =\frac{P-m}{|P-m|}
\end{eqnarray}
In the sequel, we take the previous description of the model in terms of the four game-aspiration parameters plus the effective temperature. In this way, we have fewer parameters and the additional advantage that two of them, \(\kappa_{c}\) and \(\kappa_{d}\), take values in the set \{-1,1\}. This enables us to organize the analysis by focusing on the parameter space \((\tau,\sigma,\theta)\) in three different cases:
\begin{itemize}
\item[-] Case I: \(\kappa_c=\kappa_d=1\).
\item[-] Case II: \(\kappa_c=-\kappa_d=1\).
\item[-] Case III: \(\kappa_c=\kappa_d=-1\).
\end{itemize} 
The fourth possibility, case II': \(\kappa_c=-\kappa_d=-1\), can be reduced to case II after suitable parameters change \cite{aguilar2024cooperation}. Furthermore, a possible dependence on the initial conditions due to a loss of ergodicity must also be considered.

The model's scaling property reveals how different parameters can be combined without affecting the overall dynamics.  Specifically, all games with varying values of \(S\), \(R\), \(T\), \(P\), and \(m\), but identical values of \(\sigma\), \(\tau\), \(\kappa_c\), and \(\kappa_d\) are equivalent and indistinguishable from a dynamical perspective. Nonetheless, the scaled version of the model has the drawback of diminishing our intuition of the meaning of the parameters. Therefore, after our analysis, we return to the original formulation when examining the prisoner's dilemma, using the scaled version solely for formal analysis.

\subsection{Theoretical description}

To clarify the following theoretical approach, we begin by defining three notions of state as they appear at different levels of description. A microstate is defined by \(\textbf S=(c_1,\dots,c_N)\), the state vector of all agents. A mesostate is given by \(p(\textbf S,t)\), the probability of finding the system in a given microstate  \(\textbf S\) at time \(t\). Mesostates are obtained from microstates by averaging over different realizations. Finally, when the main contribution to \(p(\textbf S,t)\) comes from microstates with similar values of \(\sum_{\mu=1}^Nc_{\mu}\), we also refer to a macrostate defined by the global cooperation fraction \(\frac{1}{N}\sum_{\mu=1}^N\mean{c_{\mu}}\). In this case, we say that \(p(\textbf S,t)\) is concentrated in microstates with similar values of \(\sum_{\mu=1}^Nc_{\mu}\). Note that the three state concepts exactly coincide when the probability function \(p\) is zero except for a single state, and provide similar information when \(p\) is narrowly distributed around one microstate.

We begin our mesoscopic description of the system with a master equation for \(p(\textbf{S}, t)\), following the standard physics approach to stochastic processes \cite{van1992stochastic, toral2014stochastic}. In the case of asynchronous strategy updates, the equation can be expressed as:
\begin{eqnarray}
  \nonumber
  \label{eq:mastersyn}
  \partial_t^a p(\textbf S,t)=\sum_{\mu\in\Sigma}&& \left[(\mathcal E_\mu^+-1)\pi_{\mu}^-p(\textbf S,t) \right. \\ && \left.+(\mathcal E_\mu^--1)\pi_{\mu}^+ p(\textbf S,t)\right],
\end{eqnarray}
where \(\partial_t^a\) is the discrete-time derivative 
\begin{equation}
  \partial_t^a p(\textbf S,t) =\frac{N}{t_0}[p(\textbf S,t+t_0/N)-p(\textbf S,t)].
\end{equation}
\(\mathcal E_\mu^+\) (resp. \(\mathcal E_\mu^-\)) is an operator that increases (decreases) the number of cooperators at \(\mu\) by one, and \(\pi_{\mu}^+=Nd_\mu p_\mu/t_0\) (resp. \(\pi_{\mu}^-=Nc_\mu p_\mu/t_0\)) is the transition rate from a defector to a cooperator (resp. cooperator to defector): 
\begin{eqnarray}
  \pi_{\mu}^+=\frac{Nd_\mu }{t_0}\left[1+\exp\left(\frac{\tau \rho_\mu+\kappa_{d} (1-\rho_\mu)}{\theta \max(1,|\tau|)}\right)\right]^{-1},&& \\
  \pi_{\mu}^-=\frac{Nc_\mu }{t_0}\left[1+\exp\left(\frac{\kappa_{c} \rho_\mu+\sigma (1-\rho_\mu)}{\theta \max(1,|\sigma|)}\right)\right]^{-1},&&
\end{eqnarray}
with
\begin{equation}
  \rho_\mu=\frac{1}{k_\mu}\sum_{\nu\in\mathcal V_\mu}c_\nu
\end{equation}
the cooperation fraction in the neighborhood of \(\mu\).

Note that, for a given \(\mu\), only two of the four terms on the right-hand side of Eq.~\eqref{eq:mastersyn} are nonzero at most. Namely, if \(c_\mu=1\), then \(\mathcal E_{\mu}^+ \pi^-_\mu p(\textbf S,t)=0\) and \(\pi^+_\mu p(\textbf S,t)=0\). Similarly, if \(c_\mu=0\) we have \(\mathcal E_{\mu}^- \pi^+_\mu p(\textbf S,t)=0\) and \(\pi^-_\mu p(\textbf S,t)=0\).

For the synchronous update, the master equation can be written as
\begin{equation}
  \partial_t^s p(\textbf S,t)=\sum_{\{\textbf S'\}}\left[\Pi(\textbf S',\textbf S)p(\textbf S',t) -\Pi(\textbf S,\textbf S')p(\textbf S,t)\right],
\end{equation}
where the sum is over all possible microstates  \(\textbf S'\) and
\begin{eqnarray}
  \nonumber  \Pi(\textbf S',\textbf S)=\frac{1}{t_0}\prod_{\mu'} \left[(c_\mu c_{\mu'}+d_\mu d_{\mu'})(1-p_{\mu'})\right. && \\ \left. +(c_\mu d_{\mu'}+d_\mu c_{\mu'})p_{\mu'}\right]. &&
\end{eqnarray}
In this case, the discrete-time derivative is
\begin{equation}
  \partial_t^s p(\textbf S,t)=\frac{1}{t_0}\left[p(\textbf S,t+t_0)-p(\textbf S,t)\right].
\end{equation}

\subsection{Approximate theory}

The mesoscopic description outlined in terms of \(p(\textbf S,t)\) is often too detailed, and solving the master equation becomes challenging, especially for synchronous updates. To simplify this, we now propose an alternative mesoscopic description for asynchronous updates, expressed in terms of \(p_\rho\) and \(q_\rho\), the probabilities of finding a cooperator and a defector, respectively, when the fraction of cooperating neighbors is  \(\rho\). Then, the probabilities \(p\) and  \(q\) of any node to be a cooperator  (cooperation fraction or cooperation density) or a defector are, respectively:
\begin{equation}
  p=\sum_{\rho}p_\rho, \hspace{0.7cm}  q=\sum_{\rho}q_\rho 
\end{equation}
that verify $p+q=1$. Moreover, a formal system of master equations for \(p_\rho\) and \(q_\rho\), with asynchronous update, read
\begin{eqnarray}
  \label{eq:seq1}
  \dot p_\rho=q_\rho\pi^+_{\rho}-p_\rho\pi^-_\rho+\sum_{\rho'}\left(p_{\rho'}\Pi^p_{\rho'\to \rho}-p_\rho\Pi^p_{\rho\to \rho'}\right)&& \\
  \label{eq:seq2}
  \dot q_\rho=p_\rho\pi^-_{\rho}-q_\rho\pi^+_\rho+\sum_{\rho'}\left(q_{\rho'}\Pi^q_{\rho'\to \rho}-q_\rho\Pi^q_{\rho\to \rho'}\right)&& 
\end{eqnarray}
where \(\pi_\rho^-\) (\(\pi_\rho^+\)) is the conditional probability rate for a cooperator (defector) with a fraction of cooperating neighbors \(\rho\) to change her strategy:
\begin{eqnarray}
  \pi_{\rho}^+=\frac{N}{t_0}\left[1+\exp\left(\frac{\tau \rho+\kappa_{d} (1-\rho)}{\theta \max(1,|\tau|)}\right)\right]^{-1},&& \\
  \pi_{\rho}^-=\frac{N}{t_0}\left[1+\exp\left(\frac{\kappa_{c} \rho+\sigma (1-\rho)}{\theta \max(1,|\sigma|)}\right)\right]^{-1}.&&  
\end{eqnarray}
In addition, \(\Pi^p_{\rho\to \rho'}\) (\(\Pi^q_{\rho\to \rho'}\)) is the conditional probability rate for the neighbors of a cooperator (defector) to change their strategies and, as a consequence, the cooperation fraction changes from \(\rho\) to \(\rho'\): When \(\rho>\rho'\) any cooperator turns a defector and  \(\rho<\rho'\) otherwise. The rates \(\Pi^p_{\rho\to \rho'}\) and \(\Pi^q_{\rho\to \rho'}\) do not only depend on \(\rho\) and \(\rho'\) but also on the neighborhood of the agents that change strategy.

The new mesoscopic description is still quite complex but can be further simplified. If we assume that the neighborhoods of two linked nodes are broadly similar, with only minor differences due to their roles, we can approximate the rates \(\Pi\) as follows:
\begin{eqnarray}
  && \Pi^p_{\rho\to \rho'} \simeq (1-\rho)\pi_{\rho'}^+ \delta_{\rho'>\rho}+\rho\pi_\rho^- \delta_{\rho'<\rho}, \\
  && \Pi^q_{\rho\to \rho'}\simeq (1-\rho)\pi_\rho^+ \delta_{\rho'>\rho}+\rho\pi_{\rho'}^- \delta_{\rho'<\rho},
\end{eqnarray}
where \(\delta_{\rho'>\rho}\)=1 for any possible transition (with the change of an agent) of the form \(\rho\ \rightarrow \rho'>\rho\) and 0 otherwise; and the same is true for the other quantities. In the case of a regular network, for example, it is \(|\rho-\rho'|=1/k\) for any possible transition, where \(k\) is the degree.

Under the previous approximations, and for a regular network of degree \(k\), the  set of equations \eqref{eq:seq1}--\eqref{eq:seq2} becomes
\begin{eqnarray}
  \nonumber \dot p_\rho&\simeq& q_\rho\pi^+_{\rho}-p_\rho\pi^-_\rho+p_{\rho+\frac{1}{k}}\left(\rho+\frac{1}{k}\right)\pi^-_{\rho+\frac{1}{k}}\\ \nonumber && +p_{\rho-\frac{1}{k}}\left(1-\rho+\frac{1}{k}\right)\pi^+_{\rho} \\ \label{eq:aproxsys1}  && -p_\rho\left[\rho \pi_\rho^-+(1-\rho)\pi_{\rho+\frac{1}{k}}^+\right], \\
  \nonumber
  \dot q_\rho&\simeq &p_\rho\pi^-_{\rho}-q_\rho\pi^+_\rho+q_{\rho+\frac{1}{k}}\left(\rho+\frac{1}{k}\right)\pi^-_{\rho}  \\ \nonumber  && +q_{\rho-\frac{1}{k}}\left(1-\rho+\frac{1}{k}\right)\pi^+_{\rho-\frac{1}{k}}\\ && -q_\rho\left[\rho\pi_{\rho-\frac{1}{k}}^-+(1-\rho)\pi_{\rho}^+\right], \label{eq:aproxsys2}
\end{eqnarray}
for \(\rho\in\{0,\frac{1}{k}, \dots,1\}\) and the "boundary condition" \(p_{-\frac{1}{k}}=p_{1+\frac{1}{k}}=0\). As desired, the approximate system \eqref{eq:aproxsys1}--\eqref{eq:aproxsys2} preserves the normalization condition $p+q$=1. Namely, summing over all value of \(\rho\) in Eq.~\eqref{eq:aproxsys1}, we have 
\begin{eqnarray}
  \label{eq:sumaeq1}
  \dot p=\sum_{\rho}\dot p_\rho = && \sum_{\rho}  \bigg\{ q_\rho\pi^+_{\rho}-p_\rho\pi^-_\rho \nonumber \\ && +p_{\rho+\frac{1}{k}}\left(\rho+\frac{1}{k}\right)\pi^-_{\rho+\frac{1}{k}} -p_\rho\rho \pi_\rho^-  \nonumber \\ \nonumber &&  +p_{\rho-\frac{1}{k}}\left(1-\rho+\frac{1}{k}\right)\pi^+_{\rho} -p_\rho(1-\rho)\pi_{\rho+\frac{1}{k}}^+ \bigg\} \\ =&&\sum_{\rho} (q_\rho\pi^+_{\rho}-p_\rho\pi^-_\rho), 
\end{eqnarray}
Similarly,
\begin{equation}
  \label{eq:sumaeq2}
  \dot q=\sum_{\rho}\dot q_\rho =\sum_{\rho}(p_\rho\pi^-_{\rho}-q_\rho\pi^+_\rho).
\end{equation}
Adding Eqs.~\eqref{eq:sumaeq1} and \eqref{eq:sumaeq2}, we finally arrive at
\begin{equation}
  \dot p+\dot q=\frac{d}{dt}\sum_{\rho}(p_\rho+q_\rho)=\sum_{\rho} (\dot p_\rho+\dot q_\rho)=0.  
\end{equation}
Eqs.\eqref{eq:sumaeq1}--\eqref{eq:sumaeq2} may have more than one steady-state solution for \(\theta=0\), which is also true for the exact and original mesoscopic description. This is because the transition rates \(\pi^\pm_\rho\) can be zero for some values of \(\rho\), as discussed later. However, for \(\theta >0\), we have \(\pi^\pm_\rho>0\), and only one steady-state solution exists.

It is important to note that the new mesoscopic theory incorporates the mean-field approximation described in Ref. \cite{aguilar2024cooperation} as a limiting case. Specifically, for a large and fully connected population, the value of \(p\) determines a unique \(\rho\). As a result, Eq.~\eqref{eq:sumaeq1} (or equivalently Eq.~\eqref{eq:sumaeq2}) simplifies to:
\begin{equation}
  \dot p=(1-p)\pi^+_{\rho}-p\pi^-_\rho,
\end{equation}
which is the mean-field equations (12) and (13) of \cite{aguilar2024cooperation}, after an appropriate change of notation. 

\section{Rational agents}
\label{sec:rationalagents}

The case of rational agents corresponds to the zero effective temperature limit \(\theta=0\): agents change their strategy when and only when their payoff is lower than their aspirations. While decision-making is completely deterministic, the entire dynamics is only deterministic for synchronous updating. In asynchronous dynamics, starting from the same initial microstate, different realizations of the dynamics choose different agents and go through different microstates in general.

In both asynchronous and synchronous dynamics, the system's evolution is specified by \(p_\mu\) in Eq.~\eqref{eq:muupdate}, the probability of a change of strategy at node \(\mu\). For \(\theta=0\), the probability \(p_\mu\) can be written as
\begin{eqnarray}
  \label{eq:cmupmu}
  && p_\mu=\Theta\left[-\kappa_{c} \rho_\mu-\sigma (1-\rho_\mu)\right], \, \text{for } c_\mu=1, \\
  \label{eq:dmupmu}
  && p_\mu=\Theta\left[-\tau \rho_\mu-\kappa_{d} (1-\rho_\mu)\right], \, \text{for } c_\mu=0,
\end{eqnarray}
with \(\Theta\) being the Heaviside step function: \(\Theta(x)=1\) for \(x> 0\) and \(\Theta(x)=0\) for \(x\le 0\). Note that we have intentionally chosen \(\Theta(0)=0\) so that a change only happens for unsatisfied agents, as already stated.
In the sequel, we will often refer to the equations ~\eqref{eq:cmupmu} and \eqref{eq:cmupmu} as \(c_\mu p_\mu\) and \(d_\mu p_\mu\), respectively.

Next, we analyze the possible stationary states of the system. To do so, we refer to those already identified in the mean-field reported in Ref.\cite{aguilar2024cooperation}. Here, we pay special attention to the role played by the network structure and the state transitions, which can occur abruptly, as demonstrated in the final part of this section.
\subsection{Consensus states}\label{sub:consensus}

\begin{figure*}[ht!]
  \centering      \includegraphics[width=\textwidth]{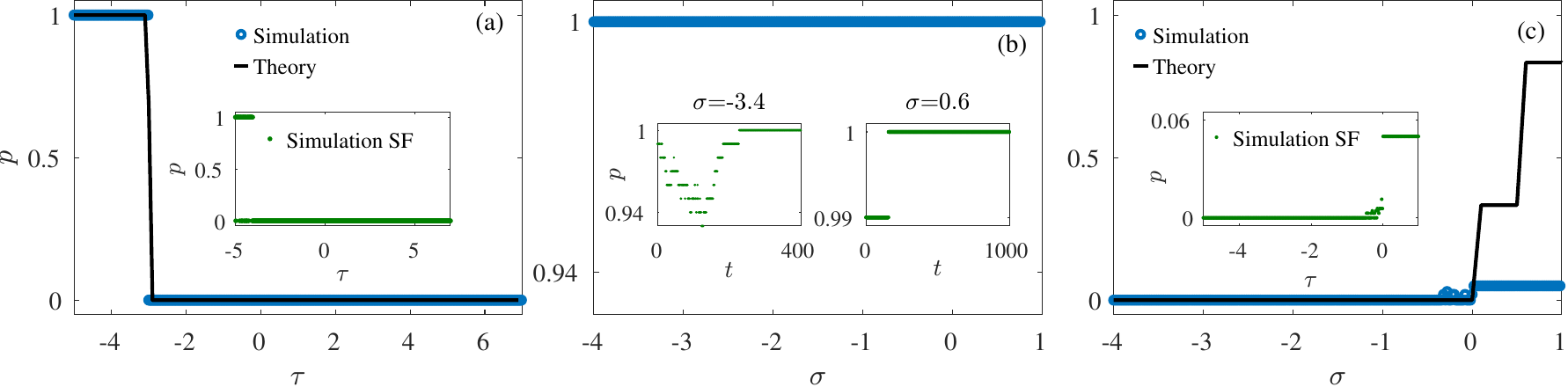}
  \caption{Monte Carlo simulations (symbols) and approximate theory (lines) for case I (\(\kappa_c=\kappa_d=1\)) and a system with $N=100$ agents following an asynchronous update. (a) The cooperation fraction $p$ as a function of $\tau$ for $\sigma=-3$ and initial state $p_0=0.95$. The main figure shows numerical simulations for the square 2-dimensional lattice, and the approximate theory. The inset corresponds to simulations in a scale-free (SF) network with mean degree $\langle k\rangle=4$. (b) The main plot shows the value of $p$ as a function of $\sigma$ for $\tau=-1$ in a SF network and an initial state in which all but one agents are cooperators, $p_0=0.99$. In the insets, we show the different dynamical behavior of this system depending on the value of $\sigma$. (c) The cooperation fraction $p$ as a function of $\sigma$ for $\tau=3$ and initial state $p_0=0.05$. The main figure shows numerical simulations for the square 2-dimensional lattice and the approximate theory. The inset corresponds to simulations in a SF network with mean degree $\langle k\rangle=4$ and $k_m=2$.}
  \label{fig:1}
\end{figure*}

The consensus states are microstates for which all agents adopt the same strategy: either all cooperate \(\textbf S=(1,\dots,1)\) or all defect \(\textbf S=(0,\dots,0)\). The accessibility of both states can be investigated by analyzing Eqs.~\eqref{eq:cmupmu} and \eqref{eq:dmupmu}.

In the case of full cooperation, since \(c_\mu=1-d_\mu=\rho_\mu=1\), Eq.~\eqref{eq:cmupmu} reduces to \( \Theta(-\kappa_c)\). Therefore, when \( \kappa_c = 1 \) (as in cases I and II), all rates become zero, indicating that cooperation consensus is an absorbing state. This will be discussed further later and represents a potential stable steady state. Conversely, when \( \kappa_c = -1 \), there is always a transition from cooperation to defection, which drives the system towards defection and prevents it from achieving full cooperation.

Let us study the stability of full cooperation when \(\kappa_c=1\) by taking all agents as cooperators except agent \(\mu\), and then analyzing the transition probability for \(\mu\) and her neighbours. For the agent \(\mu\) the transition is \(d_\mu p_\mu=\Theta(-\tau)\)  while for a neighbor \(\nu\) it is \(c_\nu p_\nu={\Theta}[-(1-1/k_\nu)-\sigma/k_\nu]\).  Therefore, for full cooperation to be stable, we require two conditions: first, \(\tau<0\) (the defector $\mu$ turns cooperator), and second  \((1-1/k_\nu)+\sigma/k_\nu\ge0\) for all her neighbors \(\nu\) (all keep cooperators). Since \(\mu\) is generic, then we need \((1-1/k)+\sigma/k\ge0\) for any degree \(k\) of the network. Note that, for \(k\) big enough, the second condition always holds, and we only require \(\tau<0\), recovering the mean-field result \cite{aguilar2024cooperation}. For the second condition, it is sufficient to ensure that it is true for the smallest degree, \(k_m\), which gives the condition \(\sigma\ge 1-k_m\).

In summary, a sufficient condition for the stability of full cooperation is
\begin{eqnarray}
  \label{eq:condcoop0}
  && \kappa_c=1 \quad  \text{(cases I and II)}, \\
  \label{eq:condcoop1}
  && \sigma\ge 1-k_m, \\
  \label{eq:condcoop2}
  && \tau<0.
\end{eqnarray}
Analogously, the sufficient condition for full defection reads 
\begin{eqnarray}
  \label{eq:conddefp0}
  && \kappa_d=1,\quad \text{(cases I and II')}, \\
  && \sigma<0, \\
  \label{eq:conddefp2}
  && \tau\ge 1-k_m.
\end{eqnarray}
For case I, \(\kappa_c=\kappa_d=1\), and \(\tau,\sigma\in(1-k_m,0)\) both full cooperation and full defection coexist.  By letting \(k_m\) go to infinity, we recover the mean-field conditions, as expected.

\begin{figure}[!t]
  \centering
\includegraphics[width=.45\textwidth]{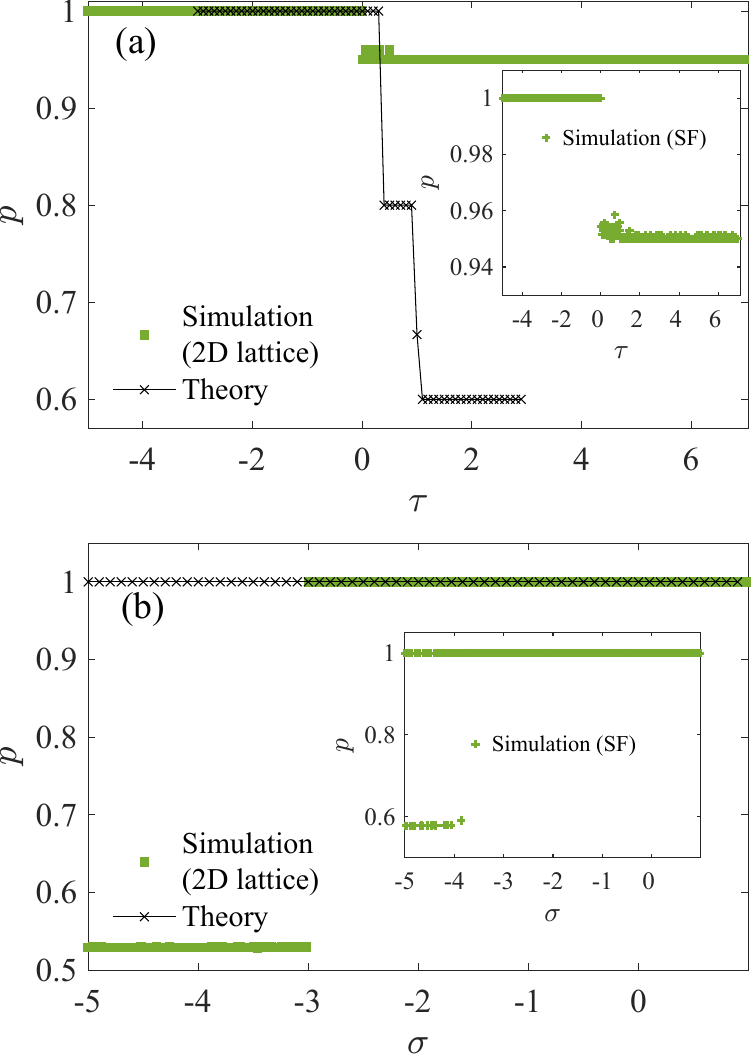} 
  \caption{Monte Carlo simulations (green symbols) and approximate theory (black crosses) for case II (\(\kappa_c=-\kappa_d=1\)) showing the cooperation fraction $p$ in a system with $N=100$ agents following an asynchronous update as a function of (a) $\tau$ for $\sigma=3$, and (b) of $\sigma$ for $\tau=-1$. In both cases, the main panels show the results for a 2-dimensional lattice and the insets for SF networks of $\langle k\rangle=4$.}
  \label{fig:2}
\end{figure}

We have conducted Monte Carlo simulations to verify the previous conditions. In Sections~\ref{sec:rationalagents} and~\ref{sec:irrationalagents}, and unless otherwise stated, the Monte Carlo simulations have been run for systems with $N=100$ agents and $t\sim 10^6$ MC steps, and results are obtained averaging observables in the last $10\%$ of a single simulation. In general, we found good agreement between the approximate theory and simulations (see Fig.~\ref{fig:1}).  However, it turned out that the conditions for full cooperation \eqref{eq:condcoop0}-\eqref{eq:condcoop2} and full defection \eqref{eq:conddefp0}--\eqref{eq:conddefp2} are sufficient but not necessary. For instance, the left inset in Fig. \ref{fig:1}(b) illustrates a system that achieves full cooperation even though the theory predicts a potential instability. Initially, the system reduces its cooperation density but eventually reaches full consensus. Additional examples of consensus states can be found in Fig. \ref{fig:2}.

\subsection{Absorbing states: full satisfaction}\label{sub:fullsatis}

The consensus states previously discussed are a subset of a broader family of microstates, the absorbing states. These are stable configurations without transitions, indicating that all agents are fully satisfied with their payoffs. 
In general, the mesoscopic characterization of consensus is not given by a probability function \( p \) that peaks at a single microstate. With the synchronous update rule, the system converges to the same absorbing state from any initial state. However, changing initial conditions or adopting asynchronous dynamics can lead to various absorbing states, even when only two consensus states exist. Not all absorbing states are reachable from every initial condition, resulting in mesostates characterized by a probability distribution \( p \) across different absorbing microstates shaped by initial conditions. 

Although the definition of absorbing states suggests a dynamic characterization, it is independent of the update rule; both synchronous and asynchronous updates can yield the same set of full-satisfaction states. The condition for an absorbing state is expressed as \( c_\mu p_\mu = d_\mu p_\mu = 0 \)  or equivalently: 
\begin{eqnarray}
  \label{eq:abscond1}
  && \sigma\ge -\frac{\rho_\mu}{1-\rho_\mu}\kappa_c, \\
  \label{eq:abscond2}
  && \tau\ge -\frac{1-\rho_\mu}{\rho_\mu}\kappa_d,
\end{eqnarray}
for all nodes \(\mu\in\Sigma\).

Conditions \eqref{eq:abscond1}-\eqref{eq:abscond2} can be verified for different values of \(\rho\), including consensus \(\rho\in\{0,1\}\), as already mentioned. Specifically, for case I (\(\kappa_c=\kappa_d=1\)), the absorbing states can be found, regardless of the value of \(\rho\), in a region of the parameter space that includes \(\sigma,\tau>0\), as seen in Fig.~\ref{fig:1}(c) and Fig.~\ref{fig:raster1}(b). {In the example illustrated in Fig.~\ref{fig:raster1}(b), the system reaches, after a short transient, an absorbing state at $p=0.36$, from which all agents are satisfied, and no change of strategy occurs. }

\subsection{Full dissatisfaction}
\begin{figure}[!h]
  \centering
    
\includegraphics[width=.45\textwidth]{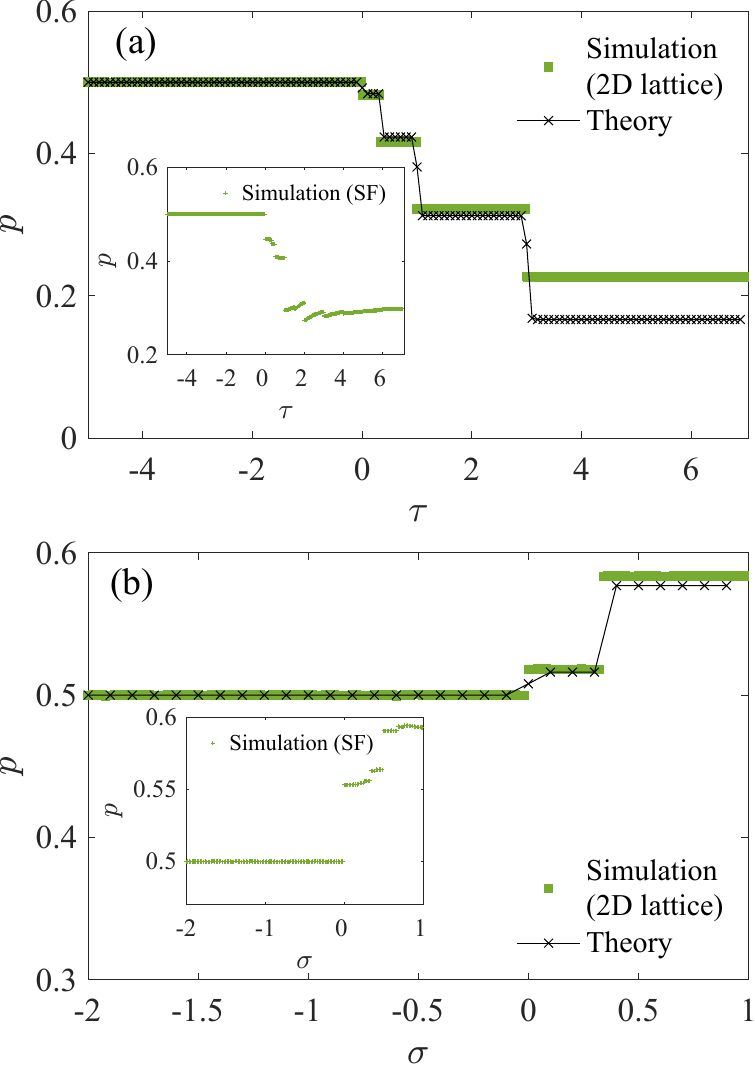}
  \caption{Monte Carlo simulations (green symbols) and approximate theory (black crosses) for case III (\(\kappa_c=\kappa_d=-1\)) showing the cooperation function in a system with $N=100$ agents following an asynchronous update as a function of (a) $\tau$ for $\sigma=-2$, and (b) of $\sigma$ for $\tau=-2$. In both cases, the main panels show the results for a 2-dimensional lattice and the insets for SF network {with $\langle k\rangle=4$}.}
  \label{fig:3}
\end{figure}

The full satisfaction scenario described in the previous section is the only mesostate in which a system can remain indefinitely. In contrast, complete dissatisfaction occurs when all nodes will change their strategy as soon as possible. 
Conditions for full dissatisfaction are found by imposing \(c_\mu p_\mu=d_\mu p_\mu=1\), that is, 
\begin{eqnarray}
  \label{eq:coexcond1}
  && \sigma<-\frac{\rho_\mu}{1-\rho_\mu}\kappa_c, \\
  \label{eq:coexcond2}
  && \tau< -\frac{1-\rho_\mu}{\rho_\mu}\kappa_d,
\end{eqnarray}
for any possible cooperation density \(\rho_\mu\) of a node's neighborhood. For case III (\(\kappa_c=\kappa_d=-1\)), the previous conditions are clearly fulfilled when \(\sigma,\tau<0\).

The observed dynamics under conditions \eqref{eq:coexcond1}-\eqref{eq:coexcond2} depend critically on the interaction network and the update rule. For asynchronous updates, the dynamics is ergodic, and all microstates are reachable. Moreover, all microstates become equiprobable in the long run and form the so-called microcanonical ensemble. Consequently, the steady probability function \(p(\textbf S)\)  symmetrically peaks around the "hypersurface" defined by \(\sum_\mu c_\mu = N/2\), with a width of the order of \(\sqrt{N}\). This allows us to identify a well-defined macrostate: the perfect coexistence state, where \(p = 1/2\).

In contrast, under the synchronous update rule, ergodicity is broken. The system becomes trapped in a cycle between two microstates: starting from a microstate \(\textbf S_1\) of full dissatisfaction, the system evolves to \(\textbf S_2\), which is the complement of \(\textbf S_1\). Since \(\textbf S_2\) also represents a state of dissatisfaction, the next state reverts to the initial state \(\textbf S_1\). In this scenario, the probability function \(p\) is \(1/2\) for both \(\textbf S_1\) and \(\textbf S_2\), while it is zero for all other microstates. As a result, a well-defined macrostate does not generally exist in this case.

If the cooperation fraction keeps around \(1/2\), the previous conditions \eqref{eq:coexcond1}-\eqref{eq:coexcond2}, become approximately 
\begin{eqnarray}
  && \sigma< -\kappa_c, \\
  && \tau< -\kappa_d,
\end{eqnarray}
which are the conditions for coexistence found at mean-field \cite{aguilar2024cooperation}. 

Examples of coexistence states are given in Fig.~\ref{fig:3} where good agreement is found for the conditions \eqref{eq:coexcond1} and \eqref{eq:coexcond2} in all reported cases, as well as with the approximate theory of Eqs.~\eqref{eq:seq1}-\eqref{eq:seq2}. Further examples are provided in Sec.~\ref{sec:prisonersdilemma}. 

\subsection{Partial satisfaction}\label{sub:partial}

\begin{figure}[!h]
    \centering
    \includegraphics[width=.48\textwidth]{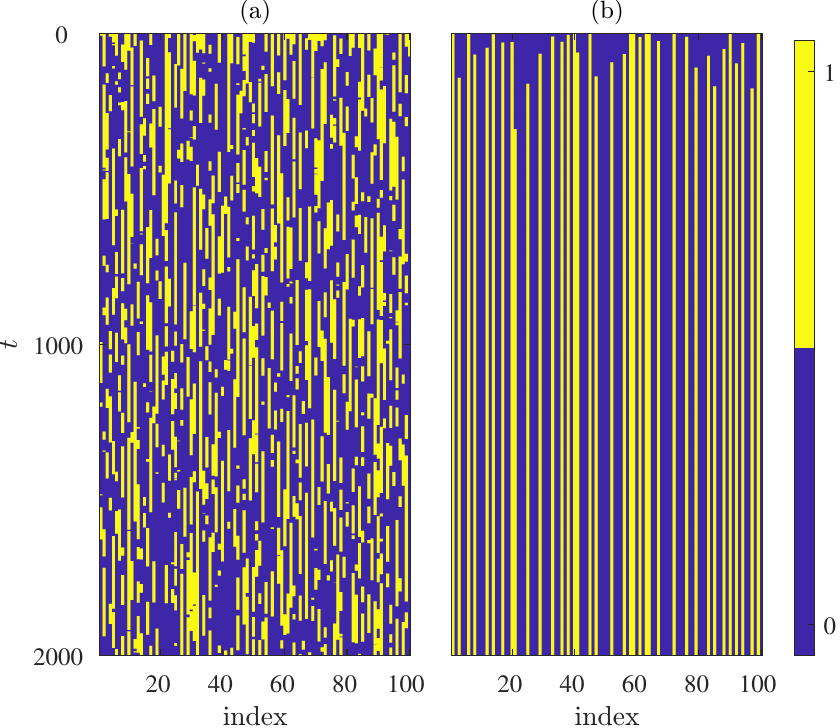}
    \caption{Raster plots showing the cooperation fraction in a two-dimensional lattice of size $N=100$ where agents update their strategies asynchronously and rationally ($\theta=0$) for (a) case II, $\sigma=-2$ and $\tau =1$ with an initial cooperation fraction $p_0=0.5$, and (b) case III $\sigma=3$, $\tau=3$, and $p_0=0.1$.
  }
    \label{fig:raster1}
\end{figure}

In this last family of mesoscopic steady states, the system navigates through microstates of partial satisfaction.  We can identify at least two situations compatible with this dynamics. The first has a subgroup of agents that remains satisfied forever with a fixed strategy in a partially absorbing state. Our simulations have not observed these dynamics since they require a specific structural configuration. In the second possible situation, all agents keep changing their strategies, not necessarily after each playing round; see Fig.~\ref{fig:raster1}(a) for an illustrative example. 

Conditions for having imperfect coexistence can be obtained by imposing satisfaction on a fraction of the population. This way, the parameters must satisfy Eqs.~\eqref{eq:abscond1}-\eqref{eq:abscond2} only for some restricted values of \(\rho_{\mu}\).

\subsection{Abrupt transitions}
\label{sub:abrupt}
In the previous sections \ref{sub:consensus}-\ref{sub:partial}, we have summarized the various stationary regimes observed in the case of rational agents. A key observation is that transitions between these regimes - when we vary the parameters \(\sigma\) and \(\tau\) - are associated with discontinuous jumps in the overall cooperation fraction \(p\). Several examples of these transitions are illustrated in Figs. \ref{fig:1} to \ref{fig:3}.

Since the dynamics for \(\theta=0\) is completely determined by the sign of the arguments of the \(\Theta\) functions in the equations ~\eqref{eq:cmupmu}--\eqref{eq:dmupmu}, a transition is expected to occur when one or both of the following conditions are met
\begin{eqnarray}
  && \kappa_{c} \rho_\mu+\sigma (1-\rho_\mu)= 0, \\
  && \tau \rho_\mu+\kappa_{d} (1-\rho_\mu) = 0, 
\end{eqnarray}
for any node \(\mu\in\Sigma\). Taking \(\rho_\mu\) as \(n/k_\mu\) with \(0\le n\le k_\mu\) a natural number, the possible transitions are located at
\begin{equation}
  \label{eq:sigman}
  \sigma_n=-\kappa_c\frac{n}{k_\mu-n}, \quad n=0,\dots, k_\mu
\end{equation}
and
\begin{equation}
  \label{eq:taun}
  \tau_n=-\kappa_d\frac{k_\mu-n}{n}, \quad n=0,\dots, k_\mu.
\end{equation}
The additional requirement for the occurrence of any of the above transitions is that the necessary value of \(\rho_\mu\) is taken at any node \(\mu\). Therefore, initial conditions can play an important role.

It is important to note that conditions \eqref{eq:sigman} and \eqref{eq:taun} are not coupled to each other. More specifically, Eq.~\eqref{eq:sigman} (Eq.~\eqref{eq:taun}) holds when a cooperator (defector) becomes satisfied or dissatisfied. In this way, by changing one parameter, each transition can induce an abrupt decrease or increase in the cooperation density, as we see explicitly in Sec.~\ref{sec:prisonersdilemma}. This property differs from what has been found in imitation-driven models as in \cite{khalil2023deterministic}, where similar but coupled conditions were found. This is because there is no direct comparison between payoffs in the aspiration-driven mechanism, unlike in imitation-driven ones. As a result, a higher number of possible transitions are expected in the first case than in the second, in general.

More examples of abrupt transitions will be provided later when we discuss the prisoner's dilemma case in Sec.~\ref{sec:prisonersdilemma}.

\section{Irrational agents}
\label{sec:irrationalagents}

\begin{figure*}[!th]
    \centering
    \includegraphics[width=\linewidth]{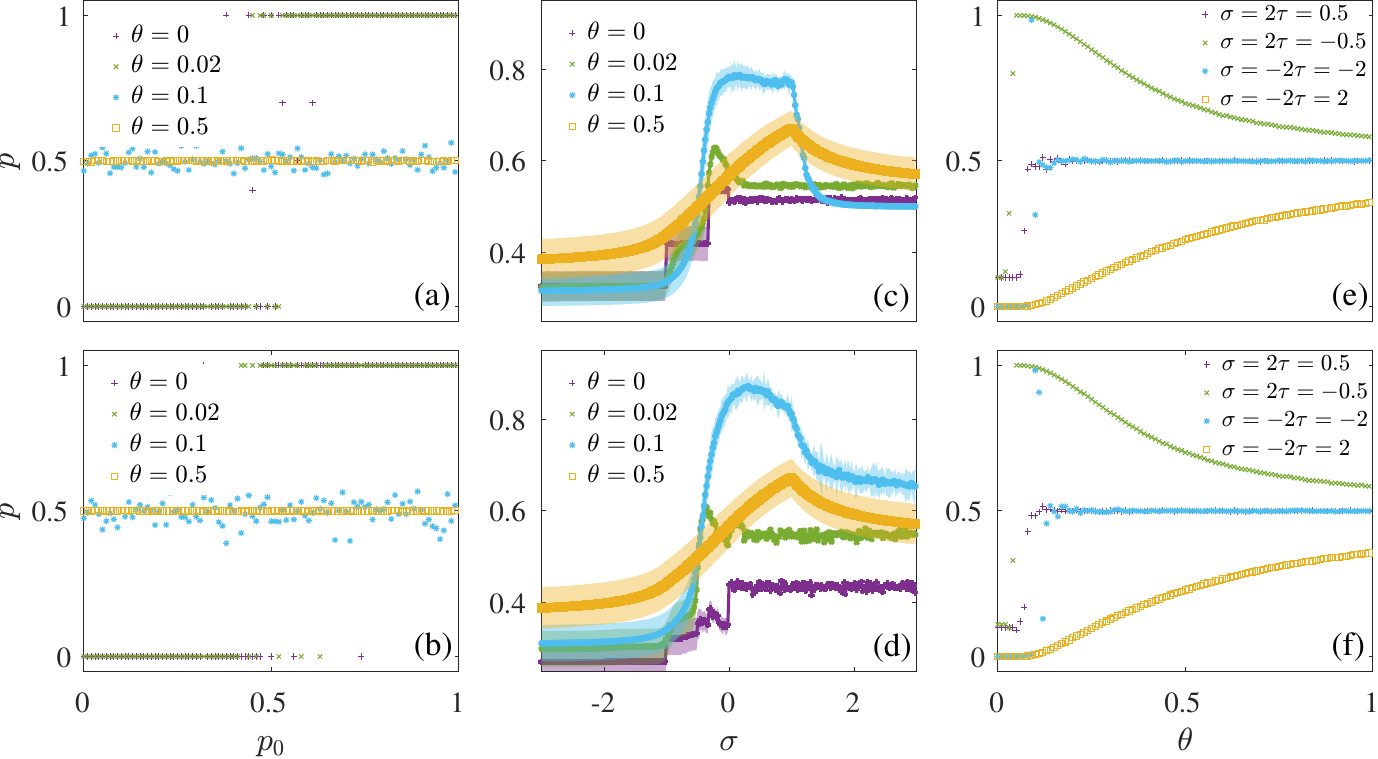}

    \caption{Transitions in the fraction of cooperators $p$ under variation of different system parameters for a system with $N=100$ agents following an asynchronous update. The upper row (a,c,e) corresponds to agents arranged in a two-dimensional lattice, while in the bottom row (b,d,f), the game is played on top of scale-free networks, and in both cases $\langle k \rangle =4$. \textbf{Left column (a,b):} Fraction of cooperators $p$ as a function of the initial fraction of cooperators $p_0$ for different values of the effective temperature $\theta$ in the case I ($\kappa_c=\kappa_d=1$) for $\tau=\sigma=-2$. \textbf{Middle column (c,d):} Fraction of cooperators $p$ as a function of $\sigma$ for different values of the effective temperature $\theta$ in the case II ($\kappa_c=-\kappa_d=1)$ and for $\tau=2$, $p_0=0.1$. Each point averages over 50 different initial conditions of microstates and network realizations.  The shadowed bands represent the standard deviation of $p$ when averaged over MC steps. Note that for low temperatures and $\sigma>0$, the system gets trapped in absorbing states, but the precise value of $p$ in which the system gets trapped depends on the specific realization of the initial state. \textbf{Right column (e,f):} Fraction of cooperators $p$ as a function of the effective temperature $\theta$, for different values of the game parameters $\sigma$ and $\tau$ in the case I and $p_0=0.1$.}
    \label{fig:similaracmedio}
\end{figure*}

When \(\theta > 0\), agents act irrationally; they may alter or maintain their strategies regardless of their payoff satisfaction. This significantly impacts the system's dynamics and potential final states. The theoretical significance of irrationality lies in its ability to render the dynamics ergodic; in other words, any microstate of the system can be accessed from any other, either directly through synchronous updating or via other microstates with asynchronous updating. 

A first consequence of ergodicity is that the system has a single steady-state probability function \(p(\textbf S)\) that is reachable from any initial condition and across various parameter values. However, the final probability \(p(\textbf S)\) can differ between synchronous and asynchronous update rules, as it will be shown in Sec.~\ref{sec:prisonersdilemma} for the Prisoner's dilemma. Furthermore, the system tends to forget the initial conditions for \(\theta>0\). Although this process can take a long time, often with a transient duration much greater than \(t_0\) (the average time an agent waits before changing her strategy), we expect that for reasonable evolution time and sufficiently low values of \(\theta\), behavior will resemble that of \(\theta = 0\).  While this phenomenon has been examined previously in the mean-field context \cite{aguilar2024cooperation}, structured interactions further slow down the dynamics. To fully understand how relaxation time is influenced by the rationality level \(\theta\), a dedicated study would be necessary, which is beyond the scope of this work. Nonetheless, significant effects tied to the interaction network are expected.

A second consequence of irrationality is that all microscopic absorbing states found for $\theta=0$ disappear, including consensus states. Even if we initially have a probability {\(p(\mathbf{S}_0)\)} concentrated in a specific microstate {\(\mathbf{S}_0\)}, i.e.
{\(p(\mathbf{S}_0)=1\)} for that microstate and 
{\(p(\mathbf{S}\ne\mathbf{S}_0)=0\)} otherwise, the system has various paths to transition to different states. As a result, \(p(\textbf S,t)\) cannot remain concentrated in a single microstate for \(t>0\). However, the system can stay close to a particular microstate, such as around complete satisfaction, if the levels of irrationality are sufficiently low.

Conversely, a full dissatisfaction regime can be sustained when $\theta>0$. In this scenario, the cumulative effects of strategy changes driven by dissatisfaction and irrational choices push the system toward $p=\frac{1}{2}$. Achieving this state of coexistence may require time and could involve significant fluctuations before reaching stability. As levels of irrationality increase with $\theta$, the transition to a steady state becomes faster. However, aside from transient effects, the degree of irrationality does not significantly influence the system's behavior during total dissatisfaction: macroscopic perfect coexistence still holds.
Additionally, stationary states characterized by partial satisfaction can arise with some degree of irrationality. As previously discussed, this irrationality inherently promotes the development of such states.

The abrupt transitions observed when agents behave rationally now become continuous, as the $p( \textbf{S})$ turns into a smooth function of the parameters when \(\theta>0\). However, as stated before, when \(\theta\) is small and over extended but realistic periods of time, the system continues to behave similarly to rational agents.

The effects of irrationality on the cooperation fraction are illustrated in Fig.~\ref{fig:similaracmedio} for some values of the parameters, two representative network types, a 2D-lattice (upper row panels) and SF networks (lower row panels), and under asynchronous dynamics. These results are compared with those in Fig.~4 of Ref.~\cite{aguilar2024cooperation} for the mean-field case. The left column of Fig.~\ref{fig:similaracmedio} shows how the cooperation fraction \(p\) depends on its initial value \(p_0\) in the case I scenario ($\kappa_c=\kappa_d=1$), and for $\tau=\sigma=-2$. For sufficiently small values of the effective temperature (\(\theta=0.02\)), the system reaches consensus under most initial conditions:  \(p\simeq 0\) for \(p_0<1/2\) and \(p\simeq 1\) for \(p_0>1/2\). Significant fluctuations between these two states are observed around \(p_0= 1/2\). This is because, for the given parameters, the consensus states are the most relevant ones for \(\theta=0\) \footnote{However, as can be seen for the lattice (see Fig.\ref{fig:similaracmedio} a), consensus states are not the only possible states for \(\theta=0\). Other absorbing states can be reached when every node \(\mu\) in the system has a cooperation fraction \(\rho_{\mu}\) in her neighborhood either greater than 2/3 or lower than 1/3, as predicted by Eqs.\eqref{eq:abscond1} and \eqref{eq:abscond2}.}. As the effective temperature increases, the system switches more frequently between consensus states. This is the case of \(\theta=0.1\), where the probability function \(p(\textbf S)\) accumulates around states with different values of the average cooperation, resulting in a no well-defined value of \(p\). Here, the value represented in Fig.~\ref{fig:similaracmedio}(a,b) is statistically significant but does not correspond to a specific macrostate. For larger \(\theta\) values, in the limit of weak selection, the cooperation fraction remains close to coexistence \(p=1/2\). No significant differences are observed between the two networks; however, there are notable discrepancies when compared to the mean-field results shown in Fig.~4 of Ref.~\cite{aguilar2024cooperation}). In a mean-field scenario, for sufficiently small \(\theta\), the cooperation fraction \(p\) reaches coexistence for \(p_0\) values near \(1/2\), while it remains at its initial value for other values of \(p_0\). The interaction structure prevents this behavior since the conditions \eqref{eq:coexcond1} and \eqref{eq:coexcond2} for coexistence are much harder to verify for the actual values of the system parameters.

The relationship between the cooperation fraction \(p\) and the rescaled parameter \(\sigma\) for different values of $\theta$ is illustrated in the middle column of Fig.~\ref{fig:similaracmedio} under the case II scenario. Here, we observe some behaviors similar to those seen in the mean-field case, specifically a non-monotonic dependence on \(\sigma\). As the temperature increases, this dependence tends to smooth out and approach \(1/2\). However, significant differences are found in the abrupt jumps in \(p\), particularly near the limit of rational agents, where the network structure plays a crucial role. These panels also reveal a noteworthy finding: a moderate value of the effective temperature, \(\theta=0.1\), maximizes the cooperation fraction on the network,  considerably exceeding the levels typically observed in mean-field. Higher temperature values attenuate this resonance and shift to higher values of  $\sigma$.  
The non-monotonic properties of \(p\) are also present when considering the dependence on \(\theta\), as shown in the right column of Fig.~\ref{fig:similaracmedio}.

\begin{figure}
    \centering
    \includegraphics[width=.45\textwidth]{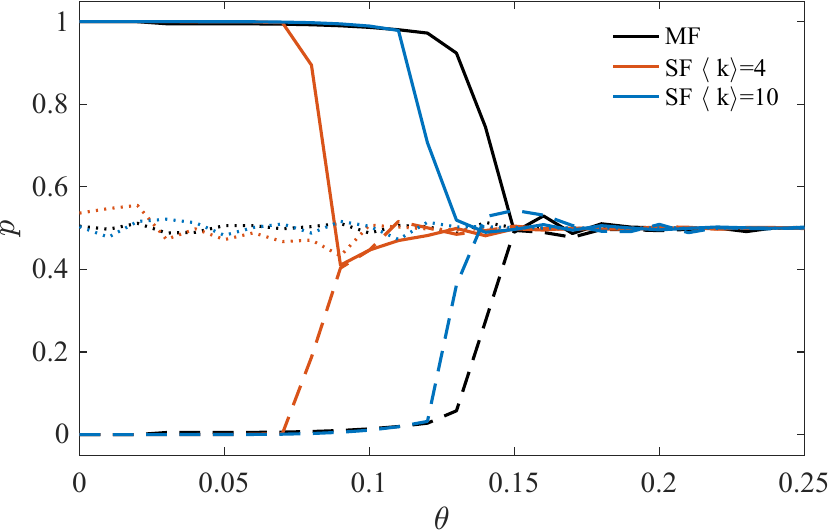}
    \caption{Steady cooperation fraction \(p\)  as a function of the effective temperature \(\theta\) for a SF network of $\langle k\rangle =4$ (in red) and $\langle k\rangle =10$ (in blue) starting from three initial cooperation fractions: $p_0=0.0$ (dashed lines), $p_0=0.5$ (dotted lines), and $p=1.0$ (continuous lines). The case of mean-field is added (in black) for comparison. The strategy update is asynchronous and game parameters are \(\tau=\sigma=-2\) and \(\kappa_c=\kappa_d=1\). Each curve averages 50 different initial conditions of microstates and 10 network realizations.
    }
    \label{fig:bifurcation}
\end{figure}

Figure~\ref{fig:bifurcation}  shows another example of how temperature can induce abrupt transitions to coexistence depending on the initial conditions. Specifically, a SF network with $\langle k\rangle=4$ (red curves) can maintain its initial state of defection (dashed line) or cooperation (solid line) for values of $\theta\lesssim 0.1$, until the cooperation level collapses into coexistence above that critical temperature. This phase transition has been reported in other systems \cite{gomez2011explosive,cai2015avalanche,lee2016hybrid,khalil2023deterministic}. Notably, the critical temperature above which a coexistence state is supported, depends on the value of the mean degree of the network $\langle k \rangle$, and grows with it toward the mean-field value, suggesting that structured interactions are more sensitive to irrational choices. 

\section{The prisoner's dilemma}
\label{sec:prisonersdilemma}

In this Section, we apply our results alongside systematic numerical simulations to examine the Prisoner's Dilemma (PD) case, setting the parameters to $S=-\frac12$, $T=\frac32$, $P=0$ and $R=1$ \cite{rapoport1989prisoner,khalil2023deterministic}. We explore how network topology, aspiration, rationality, and initial conditions influence the overall cooperation level \(p\).

\begin{figure*}[t]
  \begin{center}
    \includegraphics[width=0.75\linewidth]{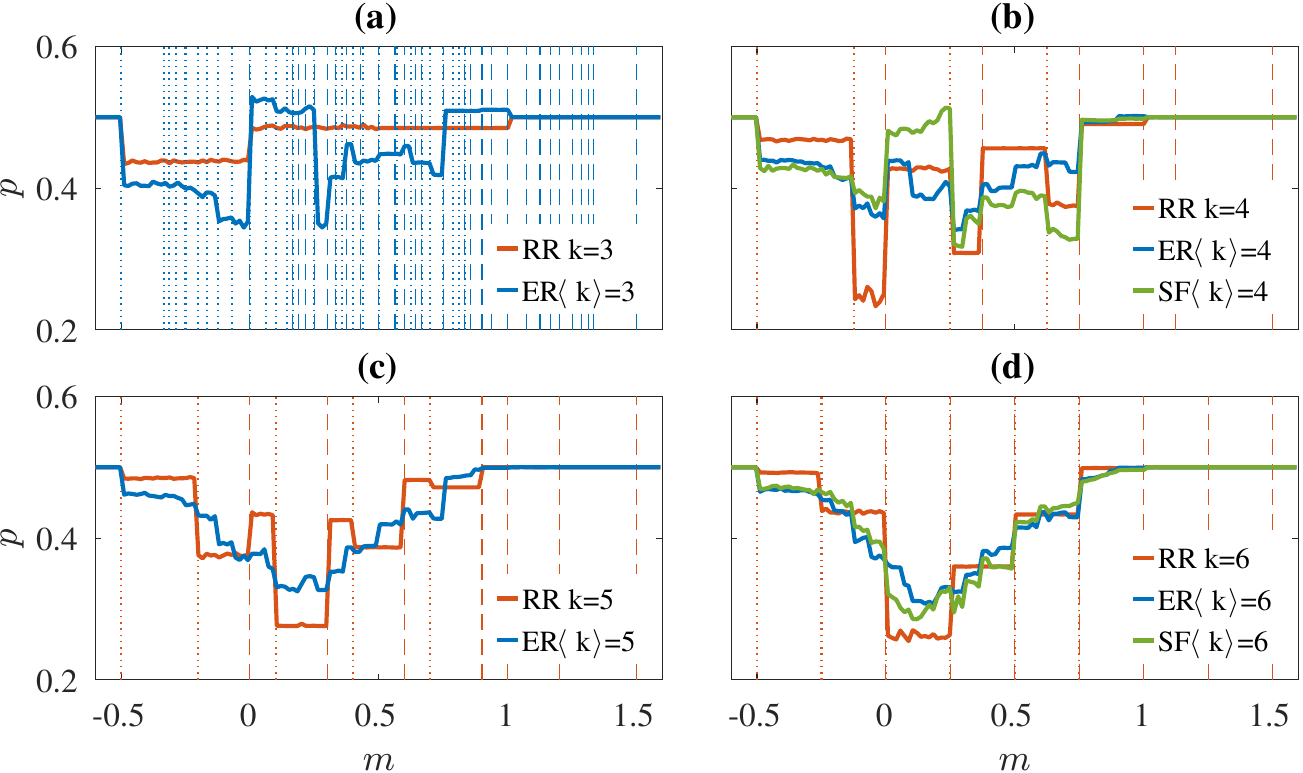}
  \end{center}
  \caption{Cooperation fraction $p$ in a Prisoner's dilemma as a function of the mood $m$ for different types of network structures and mean degrees: (a) random regular (RR) and Erd\"os-Renyi (ER) with $\langle k\rangle=3$; (b) RR, ER, and scale-free (SF) with $\langle k\rangle=4$; (c) RR and ER with $\langle k\rangle =5$; (d) RR, ER, and SF with $\langle k\rangle =6$. In all panels, dashed and dotted vertical lines correspond to values given by the critical values $m^*_{1k}(n)$  and $m^*_{0k}(n)$ respectively (Eq. \ref{eq:m*}), only for the RR network, except in panel (a) that also shows the abrupt transitions for an ER network with $\langle k\rangle=3$. In all simulations, agents' strategies are updated synchronously, and each curve is an average of over 50 initial conditions and network realizations. Rest of parameters: $N=500$, $t_{\rm total}= 5000$, $p_0=0.5$, and $\theta=0$.
  \label{fig:pd_1}}
\end{figure*}

In our simulations, we use networks of size $N=500$ of different types: Erd\"os-Rényi (ER), random regular (RR), and scale-free (SF) networks, varying the mean connectivity $\langle k\rangle\in[3,6]$, and for both synchronous and asynchronous update rules. We also make comparisons with a fully connected network~\cite{aguilar2024cooperation}. In most cases, the aspiration level varies within the range $m \in [-0.6, 1.6]$ to cover all possible behaviors for the chosen game parameters: from total satisfaction for \(m<-1/2\) to total dissatisfaction for \(m>3/2\). The effective temperature is varied within $\theta\in [0, 0.1]$, allowing us to transition from rational agents to moderately irrational ones. 
Increasing \(\theta\) further tends to remove any effects of the network topology, aspiration, and initial conditions: the system always reaches the macrostate \(p=1/2\). We also provide simulations starting from different initial conditions for the cooperation density $p_0=0.1$, $0.5$, and $0.9$. The cooperation fraction $p(m)$ and its standard deviation $\sigma_p$ are measured by averaging over $10^3 t_0$ time units after discarding a $4\times  10^3 t_0$ transient {($t_{\rm total}=5000 t_0$)}. The results are an ensemble average of over 50 network instances and initial conditions.

\subsection{Rational agents}
\subsubsection{Abrupt transitions}

As discussed in Sec. \ref{sub:abrupt}, abrupt transitions arise when a change in a parameter causes a local configuration of players (all nodes with the same strategy and the same number of cooperating neighborhoods) to change from satisfied to dissatisfied or vice versa. For the chosen values of the game parameters, the conditions for \(m\) for which abrupt transitions are expected follow immediately from Eqs.~\eqref{eq:sigman} and \eqref{eq:taun}:
\begin{equation}
  m^*_{ck}(n)=-\frac{c}{2}+\frac{3n}{2k},
    \label{eq:m*}
 \end{equation}
where  $n=0\dots,k$, and \(k\) takes values within the network's degree distribution, and $c\in\{0,1\}$. In RR networks, only a $k$ value is allowed, whereas in ER and SF networks, various connectivity degree values can exist; therefore, a wide range of possible values for $m^*_{ck}$ can yield a transition.  
 
It is important to highlight that the conditions $m^*_{1k}$ stem from an analysis of cooperation, leading to a significant drop in \(p(m)\). In contrast, $m^*_{0k}$ represents a sudden increase in \(p(m)\). When the values of $m^*_{ck}$ are degenerate, the fluctuation in \(p\) may occur in either direction.

\begin{figure}[ht!]
\centering\includegraphics[width=\linewidth]{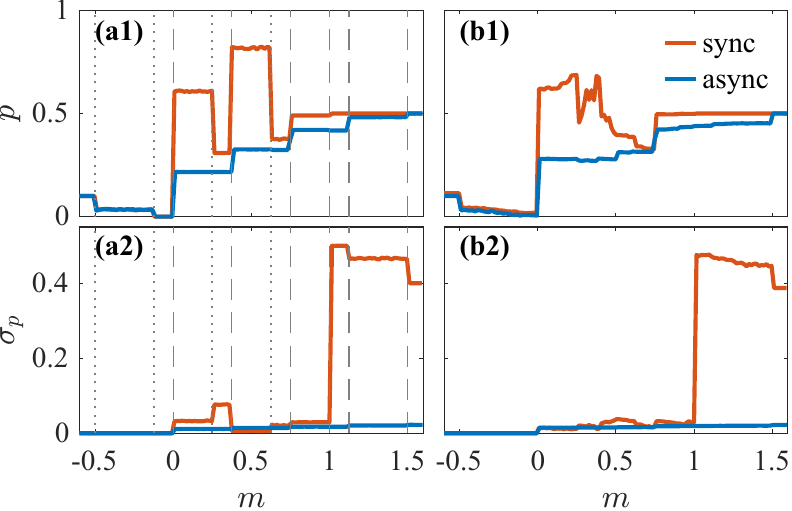} \\[0.1cm]
\includegraphics[width=\linewidth]{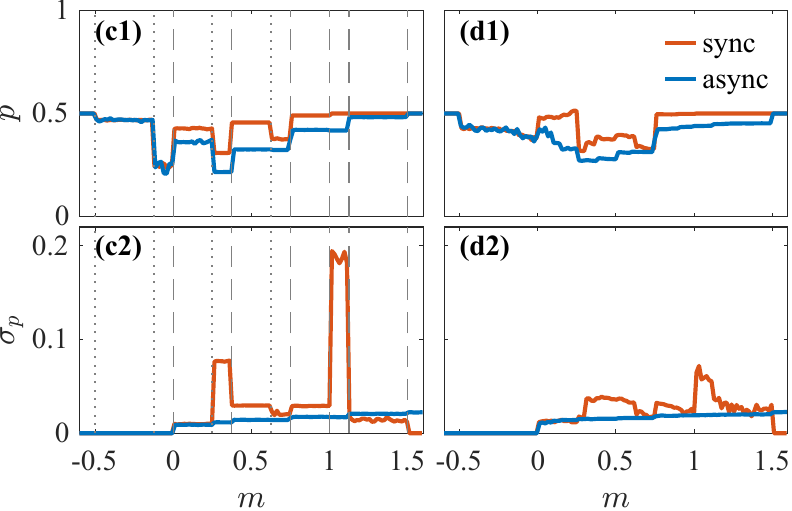}\\[0.1cm]
\includegraphics[width=\linewidth]{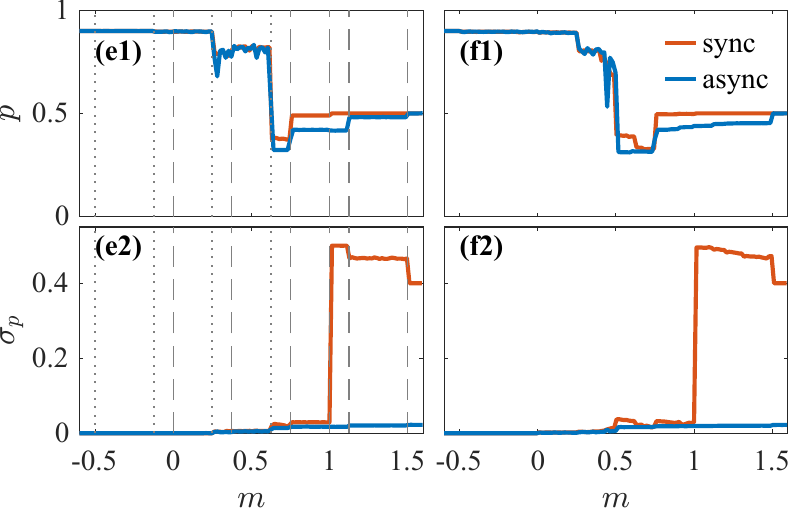}
   \caption{Comparison between synchronous and asynchronous strategy updating rules of the cooperation fraction $p$ (panels with label 1) and its fluctuation (std) $\sigma_p$ (panels with label 2) of a Prisoner's dilemma as a function of the mood $m$ for RR (left panels) and SF (right panels) networks of size $N=500$ and $\langle k\rangle=4$. The initial fraction of cooperation is  (a1-b2) $p_0=0.1$, (c1-d2) $p_0=0.5$, and (e1-f2) $p_0=0.9$. Vertical dotted and dashed lines show the abrupt transitions predicted by Eq.~\ref{eq:m*} for an RR network with $k=4$. 
   Rest of parameters: $t_{\rm total}=5000$, and $\theta=0$.
    \label{fig:pd_3}}
\end{figure}

In Fig.~\ref{fig:pd_1}, we compare the predicted $m^*_{kc}(n)$ values given by Eq.~\eqref{eq:m*} with numerical simulations for several representative cases using the synchronous update rule and starting from an initial cooperation $p_0=0.5$. The red vertical lines indicate the values of   $m^*_{1k}$ (dotted line) and  $m^*_{0k}$ (dashed line) for RR networks with $k=3$ (panel a), $k=4$ (panel b), $k=5$ (panel c), and $k=6$ (panel d). Additionally, in Fig.~\ref{fig:pd_1}(a) the predicted transition values for ER  networks with \(\mean{k} = 3 \) are marked with blue vertical lines, while the rest of the $m^*_{ck}$ values for ER and SF networks are too dense to show. For RR and SF networks, other interesting dependencies on the initial conditions and update rules are explored in Fig. \ref{fig:pd_3}. Notably, the transitions observed in the simulations align perfectly with the theoretical predictions. In the case of  ER and SF networks, the irregular shape (or less defined transitions) of \(p(m)\) is a result of multiple micro-transitions caused by complex topological interactions. 

Let us focus on the cooperation levels in Fig. \ref{fig:pd_1}. For the chosen parameters, a node \(\mu\) has a payoff per degree \(k_\mu\) in the set: 
\begin{equation}
  \label{eq:cgku}
  g_\mu/k_\mu\in\left\{-\frac{c_\mu}{2}+\frac{3n_\mu}{2k_\mu} \right\},
\end{equation}
where \(c_\mu=1\,(0)\) if $\mu$ is a cooperator (defector) and $n_\mu=0,\dots,k_\mu$. Since an agent is satisfied if $m \le g_\mu/k_\mu $, the system is stuck in an absorbing microstate of full satisfaction when $m\le m^*_{1k}(0)=-1/2$ for all values of $k$ and, therefore, the cooperation remains fixed at the initial condition \(p(m)=p_0\) as observed in Fig.~\ref{fig:pd_1} for $p_0=0.5$, and in Fig.~\ref{fig:pd_3}(a1,b1) for $p_0=0.1$, Fig.~\ref{fig:pd_3}(c1,d1) for $p_0=0.5$, and Fig.~\ref{fig:pd_3}(e1,f1) for $p_0=0.9$.

For $m \ge m^*_{1k}(0)$, close to \(m^*_{1k}(0)\), a single dissatisfaction configuration appears: a cooperator surrounded by defectors. 
Therefore, after the first round, all isolated cooperators become defectors (since the dynamics is updating all the agents synchronously). 
Note that the cooperation loss observed at the first transition at $m^*_{1k}$ coincides with the probability of finding an isolated cooperator in the initial microstate, which is approximately $\Delta p\simeq p_0\sum_{k}P_k (1-p_0)^k,$ where $P_k$ is the degree distribution. This estimation is in good agreement with the numerical results.

The former situation still holds in the interval  
\begin{equation}
  -\frac12=m^*_{1k}(0) <m\le m^*_{1k_M}(1) =-\frac12+\frac{3}{2k_M},
\end{equation}
with \(k_M\) being the network's maximum degree, which accounts for the second plateau shown in each of the four panels of Fig.~\ref{fig:pd_1}. We observe that as we increase the average degree \(\mean{k}\) of the network, both the width \(3/(2k_M)\) of the second plateau and the \(\Delta p\) reduce and eventually disappear at the mean-field limit \cite{aguilar2024cooperation}.

For \(m<0\), all defectors are satisfied, whereas all dissatisfied cooperators become defectors after the first play round (under a synchronous strategy update). This means that the transient lasts only one round, and all steady states of the system are absorbing, i.e., of full satisfaction. In contrast, for \(m>0\), some defectors initially become dissatisfied. As a result, the final state is typically not absorbing. Both isolated cooperators and defectors remain dissatisfied and exchange their roles after the first round, continuing to be dissatisfied agents.  However, {even} for a sufficiently small value of \(p_0\), the system is still expected to reach a mesostate of partial satisfaction (coexistence state, $p=0.5$). This is confirmed in Fig.~\ref{fig:pd_3} (the red curves for synchronous update), which shows \(p(m)\) for initial conditions $p_0=0.1$ (a1,b1), $0.5$ (c1,d1), and $0.9$ (e1,f1). The left panels correspond to RR networks, while the right panels correspond to SF networks. The figure also shows the fluctuations of \(p(m)\), denoted as $\sigma_p$, in the panels labeled with 2. This provides additional evidence to distinguish satisfaction from partial satisfaction; the former experiences much smaller fluctuations than the latter. As \(m\) increases further, the system consistently evolves into a state of coexistence with partial satisfaction.  

At this point, the main difference between both update rules becomes evident for intermediate values of the mood. For synchronous update (red curves), there are cases where the cooperation density reaches values higher than those observed for asynchronous update (blue curves). This is illustrated when comparing Fig.~\ref{fig:pd_3}(a1) and (c1) for RR netowkrs, or (b1) and (d1) for SF.  Synchronous dynamics allows the system to jump from low to high cooperation levels, leading to high cooperation stationary states. In contrast, asynchronous dynamics evolve {more gradually}, avoiding significant jumps in cooperation from round to round. Consequently, synchronous updates exhibit more drastic critical behavior compared to asynchronous dynamics.

Finally, for values of \(m> 3/2\), the system reaches a mesostate characterized by complete dissatisfaction, regardless of its network structure or update rule. In the case of synchronous updating, the fraction of cooperation switches from \(p_0\) to \(1-p_0\) in each round. This implies that \(p(m)=1/2\) for \(m>3/2\), with fluctuations influenced by  \(p_0\). Specifically, $\sigma_p$ is minimal when \( p_0 = \frac{1}{2} \) but increase as \( p_0 \) deviates from \( \frac{1}{2} \) (see Fig.~\ref{fig:pd_3}(a2,b2,e2,f2).

\subsection{Irrational agents}

\begin{figure}[t]
  \begin{center}
    \includegraphics[width=0.4\textwidth]{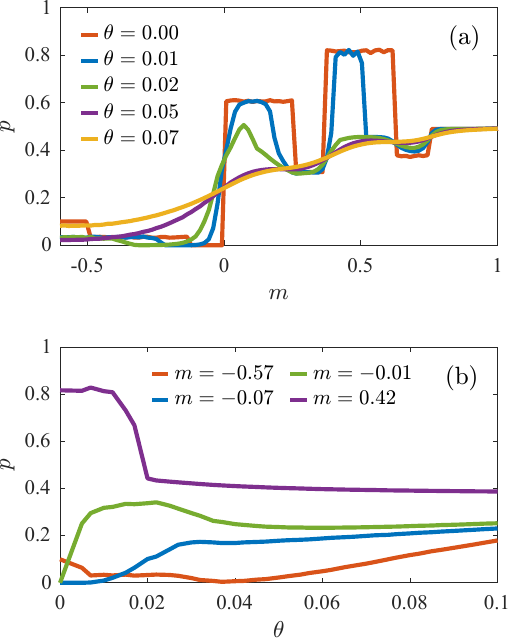}
  \end{center}
  \caption{Cooperation fraction $p$ in a Prisoner's dilemma on a RR graph with $k=4$ (a) as a function of the mood $m$ for several values of the effective temperature $\theta$, and (b) as a function of $\theta$ for several values of $m$. In all simulations, agents' strategies are updated synchronously, and each curve represents the average of over 20 initial conditions and network realizations. Rest of parameters: $N=500$, $t_{\rm total}= 5000$, $p_0=0.1$.}
  \label{fig:pd_4}
\end{figure}

For irrational agents, ergodicity introduces changes in the observed dynamics. In Fig.~\ref{fig:pd_4}, we compare the results for increasing values of the effective temperature with the rational case $\theta=0$ for the RR network with $k = 4$ under synchronous update and  $p_0=0.1$. These results can be extended to any topology and asynchronous update, provided the computational time is increased proportionally. Since Monte Carlo simulations are finite, the values displayed are not steady-state values; they have been obtained after a period longer than the transient time for $\theta=0$, which serves as the relevant timescale for comparison.

In the weak selection limit $\theta \gg 1$, all agents make decisions randomly, with a probability close to \(1/2\), independently of any topological or dynamical conditions. On the other hand, in the quasirational limit $\theta \rightarrow 0$, the finite time computational results would be indistinguishable from the rational case. However, for intermediate values of $\theta$, as shown in Fig.~\ref{fig:pd_4}a, the effects of the irrational choices depend on the value of $m$.   For \(m<-1/2\), rational agents are constrained to remain in a microstate with $p=p_0=0.1$. As \(\theta\) increases, isolated cooperators are more likely to defect, resulting in an initial decrease of $p$ with increasing \(\theta\). Since  \(p\rightarrow 1/2\) for  \(\theta\gg \), the cooperation fraction has a global minimum as we see in Fig.~\ref{fig:pd_4}(b) for $m=-0.57$.  
Around \(m=0\), the behavior of the cooperation fraction becomes more complex, exhibiting a resonant behavior for some values of $m$. If \(m\) is further increased, both dissatisfaction and temperature drive the system towards \(p=1/2\). 

\section{Conclusions}
\label{sec:conclu}

Within the framework of evolutionary game theory, we propose and analyze a model driven by aspirations, both theoretically and numerically. In this model, agents select between two strategies: cooperation and defection. They choose by comparing their payoffs to a global reference measure known as aspiration. The payoffs are influenced by a pairwise specific game, determined by four parameters, and by local configurations of strategies and neighborhood topology. Additionally, decision-making is controlled by a level of irrationality represented as an effective temperature. This factor allows us to account for additional influences that may affect strategy choice but are not explicitly included in the model. When combined, these components create a model that is both simple and comprehensive, effectively capturing essential characteristics seen in real-world applications.

A key aspect of the model is the interaction between aspiration and game parameters, which reduces the parameter space from six dimensions to a discrete set of three. This three-dimensional parameterization is the only one that is dynamically significant, as only the values derived from different rescaled parameters lead to distinct dynamic evolutions. The scaling property has been essential for systematically studying the model.

Our analysis reveals two key concepts: rationality and satisfaction. Fully rational agents focus solely on payoffs and aspirations, changing only when payoffs fall below aspirations, while satisfaction measures the number of agents with payoffs exceeding their aspirations. For rational agents and high satisfaction, the system's evolution depends on the initial conditions and the interaction network, while the specific strategy update rules have minimal impact. In contrast, high irrationality or low satisfaction leads to cooperation levels of around 1/2, regardless of network details, initial conditions, or dynamics. Between these limits, outcomes can vary significantly.

As satisfaction decreases while rationality remains high, local agent configurations become dissatisfied, leading to abrupt cooperation fraction changes. In the context of the prisoner dilemma, these transitions were analyzed by maintaining a low effective temperature and raising aspiration. Notably, the sequence of cooperation changes correlates with specific configurations of cooperators and defectors that become dissatisfied. This decoupling is novel compared to other models exhibiting abrupt transitions, such as imitation-driven models \cite{khalil2023deterministic}. Complex topologies lead to many micro transitions for intermediate aspiration values, which hide important cooperation features. We derive exact expressions for the location of abrupt changes in cooperation fractions and further develop an approximate theory that assesses how these changes depend on the full spectrum of parameters, including the location and intensity of abrupt shifts at zero effective temperature.

The cooperation fraction heavily depends on the update rule for intermediate satisfaction and high rationality, particularly in the prisoner's dilemma scenario. With asynchronous updates, cooperation progresses in two stages as aspiration rises: initially, cooperation decreases due to dissatisfied cooperators; later, it increases, approaching near-perfect coexistence. In the case of synchronous updates, a similar trend occurs with smaller aspirations. However, higher aspirations cause significant cooperation jumps when there is enough local dissatisfaction. This can lead to very large values of the cooperation fraction, even if it begins from a small initial value. This finding is noteworthy in our work and highlights an important distinction between the mean-field model \cite{aguilar2024cooperation} and other models \cite{khalil2023deterministic}.

When irrationality increases, the interaction network and aspiration's relevance decline, in agreement with previous research findings \cite{zhou_l_pre18,du_j_sr15,lim2018satisfied}. However, the transition from rationality to irrationality can be abrupt. At low, non-zero temperatures, the evolution of the cooperation mimics that of rational agents. In contrast, notable changes emerge at higher temperatures, revealing potential first-order phase transitions as seen in Fig.~\ref{fig:bifurcation}. A similar transition is explained in the mean-field case \cite{aguilar2024cooperation}, a signature of its thermodynamic nature.

As can be seen in Figs.~\ref{fig:similaracmedio}, \ref{fig:pd_1} and \ref{fig:pd_3}, for rational agents or those exhibiting low levels of irrationality, the system is not entirely insensitive to the underlying network topology. To observe the effects of the network becoming negligible, one must move to a regime of high irrationality behavior, particularly in the weak selection regime, as previously reported in Ref.~\cite{du_j_sr15}. This suggests that to model experimental setups where the network has little influence on the level of cooperation achieved \cite{grujic_pone10,grujic_srep12,gracia-lazaro_pnas12,gracia-lazaro_srep12}, it may be necessary to operate at moderately high levels of the effective temperature $\theta$. 

In conclusion, cooperation levels and satisfaction do not always align as expected; low aspirations can yield satisfaction even when cooperation is low. However, this scenario is stable only if the agents involved are rational. When aspirations are high, all agents tend to feel dissatisfied, prompting perpetual strategy changes, regardless of whether the agents are rational or not. The optimal conditions for achieving high cooperation and satisfaction occur when highly rational agents have moderate aspirations and synchronously update their strategies in a structured environment. 

Future work could extend the current model by incorporating heterogeneous aspiration levels among players, allowing for a more individualized decision-making process. A particularly interesting direction would be introducing correlations between aspiration levels and the network structure. In real-world social systems, an individual’s connectivity and social status often influence personal aspirations. Exploring how these correlated structures affect cooperation transitions and critical phenomena in social games could provide a more deeper understanding of cooperation in real-world networks.

\section*{Acknowledgments}
This research was supported by the Spanish  Ministerio de Ciencia e Innovación (Projects PID2020-113737GB-I00 and PID2023-147827NB-I00) and by the Community of Madrid and Rey Juan Carlos University through the Young Researchers program in R\&D (Grant CCASSE M2737).  M. A-J. was supported by Ministerio de Ciencia, Innovación y Universidades (Spain), Agencia Estatal de Investigación (AEI, Spain, 10.13039/501100011033), and European Regional Development Fund (ERDF, A way of making Europe) through Grant No. PID2020-112936GB-I00.

\bibliography{references}

\end{document}